\documentclass[12pt]{article}
\usepackage{graphicx}

\begin{document}

\noindent{\it\small Astronomical and Astrophysical Transactions}

\noindent{\small Vol.25, No.2, Month 2006}

\vskip 2cm

\begin{center}

\textbf{\large GOULD BELT KINEMATICS ON THE BASE OF THE OPEN
CLUSTERS AND OB-ASSOCIATIONS}\\

\bigskip

\large{Vadim~V.~Bobylev* and  Anisa~T.~Bajkova}\\

\bigskip

\normalsize

{\small Main (Pulkovo) Astronomical Observatory of the Russian
Academy of Sciences, Pulkovskoye Chaussee 65/1, St-Petersburg
196140,  Russia\\}

\bigskip

*Corresponding author. Email: vbobylev@gao.spb.ru\\

\vskip 1cm

{\it\small (Received 29 June 2006)}

\vskip 1 cm

\end{center}

\bigskip

\small
\noindent The kinematic parameters of the Gould Belt are
redetermined using modern data on the motion of nearby young
($\log t<7.91$) open clusters, OB-associations and moving stellar
groups. The modelling carried out shows that the residual
velocities achieve the maximal values equal to $-4$~km s$^{-1}$
for rotation (in the direction of the Galactic rotation) and
$+4$~km s$^{-1}$ for expansion,  when the distance from the
kinematic centre is about 300 pc. We assumed the following
parameters of the Gould Belt centre: $l_\circ=128^\circ$ and
$R_\circ=150$~pc. It is shown that the whole structure of the
Gould Belt is moving relative to the local standard of rest with a
velocity of $10.7\pm0.7$~km s$^{-1}$ in the direction towards
$l=274\pm4^\circ$, $b=-1\pm3^\circ$. Using the rotation velocity,
we obtained the virial estimation of the Gould Belt mass as
$1.5\times 10^6 M_\odot$.\\

\bigskip


\vskip 1 cm

\normalsize
\noindent The Gould Belt is a system of nearby OB-associations
[1], HI clouds [2], H$_2$ molecular cloud complexes [3]. At
present, there exist several  scenarios for the formation of the
Gould Belt. According to the first scenario, the Gould Belt is
formed as a result of the collision of high-speed neutral hydrogen
clouds with the Galactic disk [4]. According to the second
scenario, the Gould Belt is formed owing to explosion of
supernovae [2, 5]. According to the third scenario [6], the
formation of the Gould Belt is a stage of kinematic evolution of
the Local Stellar System. Linblad [7] suggested a model of proper
rotation and expansion of the Gould Belt which takes into account
the inclination of the disk to the Galactic plane ($20^\circ$). In
the work of Bobylev [8], the well-known Linblad approach is
developed for the case when star heliocentric distances are
determined exactly.

The goal of this work is redetermination of kinematic parameters
of the Gould Belt structure using modern data on open clusters.
The advantage of such an approach is that the estimations of  ages
and distances of open clusters  are more reliable than those of
individual stars.

The catalogue of 652 open clusters hereafter we denote as COCD
[9,10]. Our working list is composed primarily from the COCD young
open clusters with age $\log t<7.9$, denoted in the paper of
Piskunov et al. [10] as OCC1. In the COCD catalogue, random errors
of mean proper motions of the clusters are small because of the
use of the combined catalogue ASCC-2.5 [9]. The radial velocities
of the COCD open clusters are based on the catalogue published by
Barbier-Brossat and Figon [11]. We use more comprehensive star
radial velocities from the Orion Spiral Arm Catalogue (OSACA)
[12], which combines a number of large catalogues of radial
velocities including those of Duflot {\it et al.} [13],
Barbier-Brossat and Figon [11], Nordstr\"om {\it et al.}[14] and
Famay {\it et al.}[15]. In comparison with the work in [3], we
made essential supplements for stars of the  Sco-Cen complex using
the list compiled by de Zeeuw {\it et al.} [1] and the OSACA. We
considered the open cluster Chamaeleontis. We used modern data for
25 stars of cluster TWA as well.
Our approach to kinematic analysis is based on the following
equations [8]:
$$
\displaylines{\hfill
  V_r= -u_{\odot}\cos b\cos l-v_{\odot}\cos b\sin l-w_{\odot}\sin b+\cos^2 b k_\circ r+
\hfill\llap(1)\cr\hfill
  +(R-R_\circ)(r\cos b-R_\circ\cos (l-l_\circ))\cos b  k'_\circ-
 R_\circ (R-R_\circ)\sin (l-l_\circ) \cos b \omega'_\circ,\hfill\cr
\hfill 4.74 r \mu_l\cos b=u_{\odot}\sin l-v_{\odot}\cos l -
\hfill\llap(2)\cr\hfill
  -(R-R_\circ)(R_\circ\cos (l-l_\circ)-r\cos b)\omega'_\circ
 +r\cos b \omega_\circ+R_\circ(R-R_\circ)\sin (l-l_\circ)) k'_\circ,\hfill
 \cr
\hfill 4.74 r \mu_b=
 u_{\odot}\cos l\sin b+v_{\odot}\sin l \sin b-w_{\odot}\cos b-\cos b\sin b k_\circ r-
  \hfill\llap(3) \cr\hfill
 -(R-R_\circ)(r\cos b-R_\circ\cos (l-l_\circ))\sin b k'_\circ
 +R_\circ(R-R_\circ)\sin (l-l_\circ)\sin b\omega'_\circ.\hfill
 }
$$
Here  $r=1/\pi$ is the heliocentric distance of a star, $R_\circ$
is the distance from the Sun to kinematic centre,$l_\circ$ is the
direction to the centre, $u_{\odot},v_{\odot},w_{\odot}$ are the
components of the solar peculiar velocity. The components
$\mu_l\cos b$ and $\mu_b$ are expressed in milliarcseconds per
year (mas yr$^{-1}$), the radial velocity $V_r$ in kilometers per
second (km s$^{-1}$), the parallax $\pi$ in milliarcseconds and
the distances $R$, $R_\circ$ and $r$ in kiloparsecs (kpc). The
quantity $\omega_\circ$ is the angular velocity of rotation,
$k_\circ$ is the radial expansion/contraction velocity of a star
system at the distance $R_\circ$ and parameters $\omega'_\circ$
and $k'_\circ$ are the corresponding derivatives. The distance $R$
from a star to the centre is evaluated as $R^2=r^2\cos^2
b-2R_\circ r\cos b\cos (l-l_\circ)+R^2_\circ$.  The system of
conditional equations (1)--(3) contains seven unknowns which we
determine using the minimum least-squares technique. It is assumed
that the left-hand sides of the equations are free from the
Galactic differential rotation.
\begin{figure}[t]
{
\begin{center}
 \includegraphics[width=80mm]{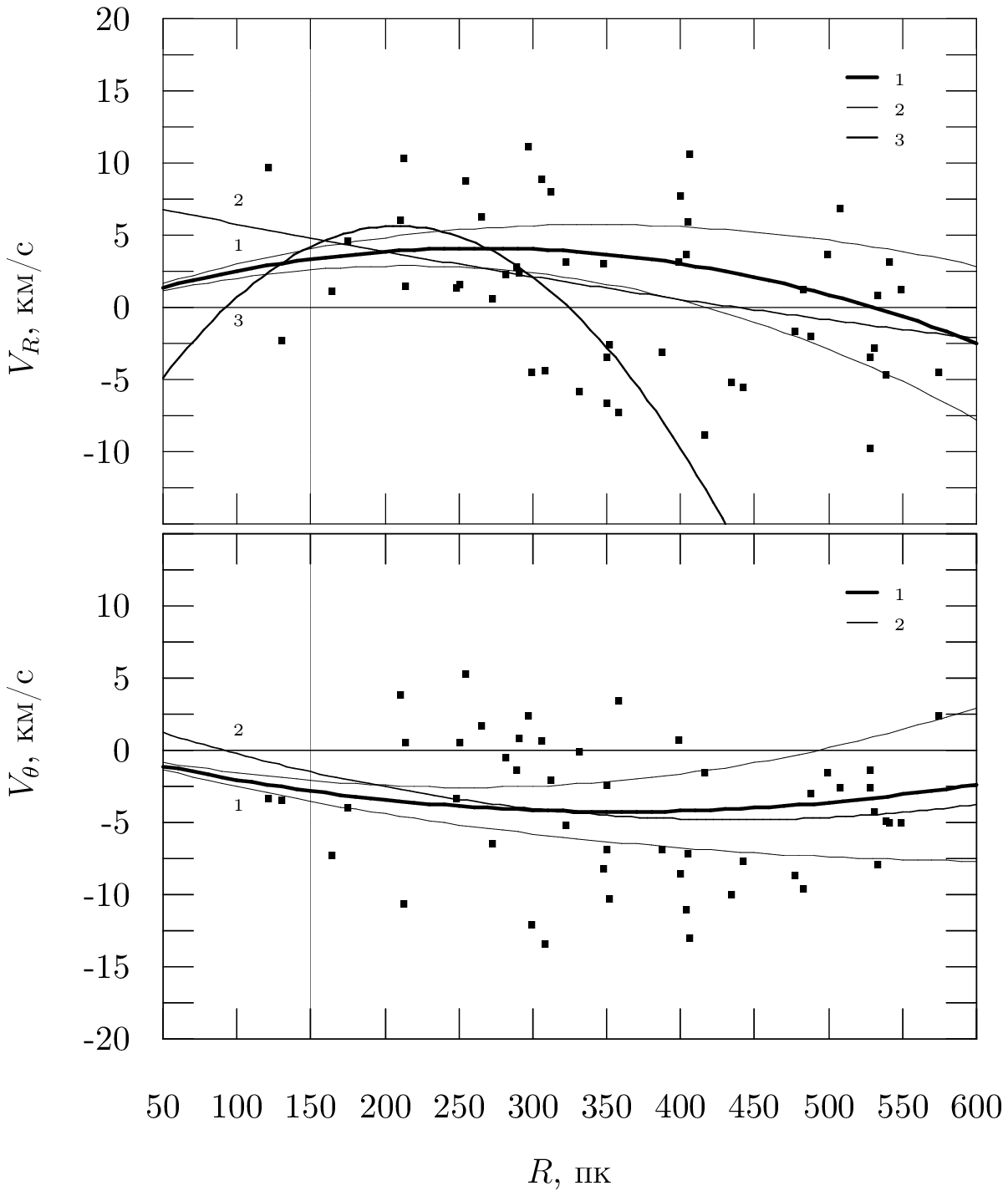}
\end{center}
} {\bf Figure~1.} {\small Residual velocities $V_R$ (top) and
$V_\theta$ (bottom) versus distance $R$  from the kinematic centre
$l_\circ=128^\circ$, $R_\circ=150$~pc; the vertical line indicates
the value of $R_\circ$.}
\end{figure}

The second method of kinematic analysis is based on linear
residual velocities which we approximate as:
$$
\displaylines{\hfill
  V_i= V_{\circ, i}+a_i (R-R_\circ)+ b_i (R-R_\circ)^2, \hfill\llap(4)
 }
$$
where $i=(R,\theta$), and the residual velocity $V_R$ of expansion
and the residual velocity $V_\theta$ of rotation  must be obtained
by decomposition of the spiral velocity components U and V into
radial and tangential parts using the known centre parameters
$l_\circ$ and $R_\circ$.
The results of the analysis are presented in figures 1 and 2. We
can conclude that it is shown in our work that the Gould Belt
structure takes part in several motions. Firstly, the whole
complex, free from common Galactic rotation, is moving relative to
the local standard of rest [16] with a velocity of $10.7\pm0.7$~km
s$^{-1}$ in the direction towards $l=274\pm4^\circ$,
$b=-1\pm3^\circ$. Secondly, residual rotation and expansion of the
whole system exist. We used the following parameters of the
kinematic centre: $l_\circ=128^\circ$ and $R_\circ=150$ pc [8].
The modelling carried out shows that the residual velocities are
reliable in a small region near $R\approx R_\circ$ and are equal
to $-2.8\pm0.7$~km s$^{-1}$ for rotation (in the direction of the
Galactic rotation), and  to $+3.3\pm0.7$ km s$^{-1}$ for
expansion. The maximal values are achieved when the distance from
the kinematic centre is about 300 pc and are equal to
$-4.3\pm1.9$~km/c for rotation and to $+4.1\pm1.4$~km s$^{-1}$ for
expansion. Using the rotation velocity, we obtained the virial
estimation of the Gould Belt mass as $1.5\times 10^6 M_\odot$.

\begin{figure}[t]
{
\begin{center}
  \includegraphics[width=80mm]{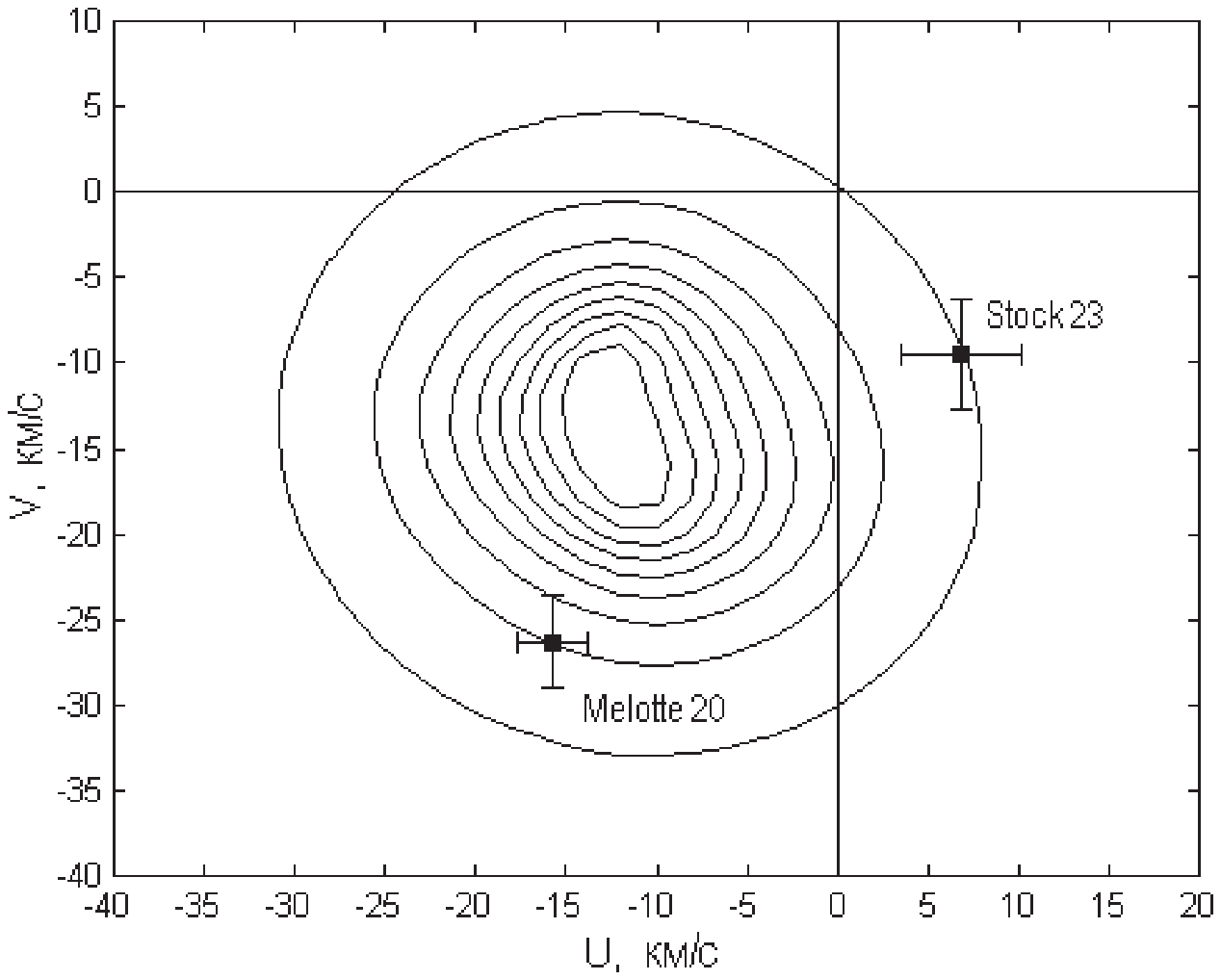}
\end{center}
} \centerline{{\bf Figure~2.}
 {\small Smoothed distribution of $UV$-residual velocities.}}
\end{figure}

\bigskip

\noindent{\bf Acknowledgement}

\bigskip

\noindent This work is supported by the Russian Fund for Basic
Research (grant No~05--02--17047).

\end{document}